\documentclass[12pt,draftcls,onecolumn]{IEEEtran}                       
\IEEEoverridecommandlockouts                              
\usepackage{graphics} 
\usepackage{epsfig} 
\usepackage{mathptmx} 
\usepackage{times} 
\usepackage{amsmath} 
\usepackage{amssymb}  
\usepackage{floatrow}
\usepackage{multirow}
\usepackage{algorithm}
\usepackage{algpseudocode}
\usepackage{color}
\newtheorem{theorem}{\indent Theorem}

\newtheorem{proposition}[theorem]{\indent Proposition}
\newtheorem{observation}[theorem]{\indent Observation}

\title{\LARGE \bf
Real-time Information, Uncertainty and Quantum Feedback Control}
\author{Bo Qi, Daoyi~Dong, Chunlin~Chen, Lijun~Liu, Zairong~Xi
\thanks{This work was supported by the National Natural Science
Foundation of China (Nos. 61374092, 61227902, 61134008 and 61004049), and by the Australian Research Council's Discovery Projects funding scheme
under Project DP130101658.}
\thanks{B. Qi is with the Key Laboratory of Systems and Control, Academy of
Mathematics and Systems Science, Chinese Academy of Sciences,
Beijing 100190, China (email: qibo@amss.ac.cn).}
\thanks{D. Dong is with the School of Engineering and Information
Technology, University of New South Wales, Canberra, ACT 2600, Australia (email:
daoyidong@gmail.com).}
\thanks{C. Chen is with the Department of Control and System Engineering, Nanjing
University, Nanjing 210093, China
(email: clchen@nju.edu.cn).}
\thanks{L. Liu is with  the Key Laboratory of Systems and Control, Academy of
Mathematics and Systems Science, Chinese Academy of Sciences,
Beijing 100190, China (email:
lljcelia@126.com).}
\thanks{Z. Xi is with the Key Laboratory of Systems and Control, Academy of
Mathematics and Systems Science, Chinese Academy of Sciences,
Beijing 100190, China (email:
zrxi@iss.ac.cn).}
}

\begin{document}

\maketitle
\begin{abstract}
Feedback is the core concept in cybernetics and its effective use has made great success in but not limited to the fields of engineering, biology, and computer science. When feedback is used to quantum systems, two major types of feedback control protocols including coherent feedback control (CFC) and measurement-based feedback control (MFC) have been developed. In this paper, we compare the two types of quantum feedback control protocols by focusing on the real-time information used in the feedback loop and the capability in dealing with parameter uncertainty. An equivalent relationship is established between quantum CFC and non-selective quantum MFC in the form of operator-sum representation. Using several examples of quantum feedback control, we show that quantum MFC can theoretically achieve better performance than quantum CFC in stabilizing a quantum state and dealing with Hamiltonian parameter uncertainty. The results enrich understanding of the relative advantages between quantum MFC and quantum CFC, and can provide useful information in choosing suitable feedback protocols for quantum systems.
\end{abstract}


\section{Introduction}\label{Sec1}
As pointed out by Norbert Wiener, feedback widely exists in machines and animals, and it actually exists in all purposeful behaviors \cite{wiener}. Feedback is the core concept in cybernetics, and its effective use has made great success in but not limited to the fields of engineering, biology, and computer science.  The essence of feedback lies in its capability in dealing with various kinds of uncertainties (e.g., uncertainties in initial conditions, parameter uncertainties, external disturbances, or all) \cite{guo,L-Guo,lichanying,huang}. Most robust control results were developed in the form of feedback control \cite{petersen-tempo}. Some fundamental results have also been presented in concerning the maximum capability of feedback in dealing with uncertainties of classical (non-quantum) systems \cite{guo,L-Guo,lichanying,huang}.

With the rapid development of emerging quantum technology, feedback control has been employed to design control laws for manipulating quantum systems in recent years \cite{judson,dong2010,wisemanbook,sayrin,vijay,ticozzi,kurt-wang}. The early development in quantum feedback control focused on measurement-based feedback control (MFC) schemes where measurement outcomes are used for the design of feedback controllers \cite{wiseman93,jacobs2007,wiseman02,Ahn-Doherty,Wiseman-Doherty,
handel-stockton-mabuchi,Mirrahimi-Handel-2007}. One significant difference in feedback for classical and quantum systems is that the measurement can play different roles in the control process. For a classical system, the backaction effect of a measurement on the system can be neglected in principle. That is, the measurement does not change the system state itself. However, when we make a measurement on a quantum system, the measurement outcome is stochastic and the measurement process usually changes the system state itself (i.e., quantum state collapse) \cite{nilsen}. The backaction effect of quantum measurement cannot be ignored for quantum systems. Therefore,  we have to deal with the additional but inherent measurement-induced uncertainty in quantum measurement-based feedback control \cite{qi-guo}. Moreover, the fast time scale in quantum systems and time delay in the feedback loop make it difficult to implement quantum MFC for practical quantum systems.

Considering the difficulty of quantum MFC, another quantum feedback control strategy, i.e., quantum coherent feedback control, has been developed \cite{wiseman-milburn,lloyd,James--Nurdin-Petersen,Mabuchi,nurdin,ZhangJames,hamerly,shaiju,iida,zhangjing}. In coherent feedback control (CFC), no explicit measurement
is involved and the controlled system is coupled to another quantum system (controller) in such a way that a quantum information flow occurs between the two systems. Quantum CFC has been widely investigated for quantum optical systems \cite{wiseman-milburn,James--Nurdin-Petersen,Mabuchi,nurdin,iida}, and shows significant advantages over quantum MFC in practical implementation due to the fact that the feedback controller has similar time scale to the system plant.

In order to better understand and perform quantum feedback control, a natural question is to compare the relative merits between these two types of quantum feedback control protocols. In \cite{kurt-wang}, a comparison was presented and the authors showed the advantages of  quantum CFC over measurement-based feedback control by analyzing specific performance that the two types of feedback control can achieve. However, the definition of quantum MFC in \cite{kurt-wang} was restricted. In this paper, the aim is to compare quantum MFC with quantum CFC by focusing on the real-time (estimated) state information used in the feedback loop and the capability in dealing with Hamiltonian parameter uncertainty. In quantum MFC \cite{Doherty-et-al,Handel-Stockton2005,jacobs06,qibo1,jzhang2010,Bolognani2010,amini,qi-pan-guo,qi},
we repeatedly measure the system and apply control operations based on available information including measurement information as well as prior information. The use of the available real-time information makes it possible to achieve better performance than quantum CFC in dealing with uncertainties.

The paper is organized as follows. In Section II, we compare the advantage of feedback control over open-loop control by an example. The aim is to emphasize the fact that the essence of feedback lies in its capability in dealing with uncertainties. In Section III, we present a quantum CFC protocol and a quantum MFC protocol in the form of operator-sum representation. We demonstrate that  in principle, there always exists an equivalent non-selective quantum MFC protocol for each quantum CFC protocol in the sense of operator-sum representation. Section IV presents a specific quantum control problem to show that quantum MFC using real-time information can achieve better performance than non-selective quantum MFC. The aim is to show that quantum MFC may have advantages over quantum CFC if the real-time information is used to design a feedback control law. Using an example, we further show quantum MFC can achieve better performance than quantum CFC in dealing with Hamiltonian parameter uncertainty in Section V. Concluding remarks are given in Section VI.

\section{Advantages of Feedback Control over Open-loop Control}\label{second}
In order to control a system, two typical classes of control strategies including open-loop control and feedback control can be implemented. In an open-loop control strategy, all we can use for designing a control law is the prior information of the system. Generally, an open-loop control strategy is simpler and easier to be implemented than the relevant feedback control strategy. For a classical (non-quantum) control system, if all information is perfect (i.e., there is no uncertainty or imperfection with the system to be controlled), then there always exists a corresponding  equivalent open-loop control protocol for each feedback control protocol. In principle, a given feedback control law that is designed based on the system state or the output of the system may be considered as a functional depending only upon the time $t$ and the initial condition. That is, there always exists an implicit equivalent open-loop control law for a given feedback control law. Hence, if all information is perfect, it may be not necessary to employ a feedback control strategy to control the system under consideration.

However, the existence of uncertainties or imprecision (e.g., external disturbance, noise in control signals) is unavoidable when our concern is the effect of the control on practical systems. The difference between a practical system and its built model can also be regarded as uncertainties or imperfection. In general, the open-loop control strategy is only valid for an ideal system model. When dealing with uncertainties or imperfection in a control system, the advantage of feedback control over open-loop control can be well shown. The advantage and essence of feedback has been widely investigated for classical control systems (see, e.g., \cite{guo,L-Guo,lichanying,huang}). Here, we demonstrate this fact with a simple example.

{\bf Example 1.} Consider the linear system
\begin{equation*}
\dot{x}=x+u,\   \   x(0)=x_0,
\end{equation*}
where $x$ and $u$ denote the system state and the control input, respectively. The aim is to stabilize the system state to $x=0$ asymptotically.

The control objective can be achieved by employing the following feedback control law $$u=-2x(t).$$ If all information is perfect (e.g., the initial state information $x(0)=x_0$ is accurate), then the open-loop control law $u=-2{e}^{-t}x_0$ is equivalent to the feedback control law $u=-2x(t)$.
However, once there is an error in the initial state $x(0)$ (e.g., the measured initial state  $\bar{x}_0\neq x_0$), feedback control can make significant difference from open-loop control. We may verify that when the open-loop control law $u=-2e^{-t}\bar{x}_0$ is applied, the state $x(t)$ will evolve as $$e^{-t}\bar{x}_0+e^{t}(x_0-\bar{x}_0).$$ Hence, the system state will exponentially diverge once $\bar{x}_0\neq x_0$. In contrast, the system state will exponentially converge to 0 with the feedback control law $u=-2x(t)$.

The above example shows the advantage of feedback control over open-loop control in dealing with uncertainties. Actually, the advantage has been well understood for classical control systems. For example, most robust control results have been developed in the form of feedback control \cite{petersen-tempo}. The capability of feedback mechanism in dealing with uncertainties has been extensively investigated \cite{guo,L-Guo,lichanying,huang}. Although feedback control has also been applied to manipulate some specific quantum systems, it is essentially different from feedback control of classical systems due to the unique characteristics of quantum systems (e.g., the measurement backaction and fast time scale). Two types of quantum feedback control including measurement-based feedback control (MFC) and coherent feedback control (CFC) have been developed for manipulating quantum systems. In this paper, we will compare quantum MFC and quantum CFC by focusing on the capability in dealing with uncertainties in quantum systems. It is worth pointing out that even there is no uncertainty or imperfection in a quantum system, a quantum MFC strategy is also not equivalent to its relevant open-loop control strategy \cite{qi-guo}. This is due to the fact that quantum measurement will induce the quantum state collapse and can be regarded as a control means itself. In the following, we will first present quantum CFC and quantum MFC in the form of operator-sum representation and then consider two classes of uncertainties in quantum systems: uncertainties in the initial state and in Hamiltonian parameters.

\section{Quantum CFC and Quantum MFC}\label{quantumcfcsection}
In this section, we establish an equivalent relationship between quantum CFC and non-selective quantum MFC in the form of operator-sum representation following a brief introduction to quantum systems.

\subsection{Quantum systems}
Associated to a quantum system is a complex vector space with inner product (i.e., a Hilbert space) known as the state space of the system. The state of the quantum system (e.g., a spin system, or a photon) can be described by a density operator $\rho$ in its underlying Hilbert space.
The density operator $\rho$ is a positive operator with trace one (i.e., Tr$(\rho)=1$). If a density operator satisfies Tr$(\rho^2)<1$, we call the quantum state $\rho$  a mixed state. Otherwise, we call it a pure state.  For a pure state (i.e., Tr$(\rho^2)=1$), if we denote the eigenvector of $\rho$ with eigenvalue being 1 as $|\varphi\rangle$, the state $\rho$ can be denoted in terms of the state vector $|\varphi\rangle$ as $$\rho=|\varphi\rangle(|\varphi\rangle)^{\dag}\triangleq|\varphi\rangle\langle\varphi|.$$
Here $\dag$ denotes the Hermitian adjoint. Generally, the state of an open quantum system  needs to be described as a mixed state.

The evolution of a closed quantum system is described by a unitary transformation. For example, the state $\rho_0$ of a quantum system at time $t_0$ is related to the state $\rho_1$ of the system at time $t_1$ by a unitary operator $U$ satisfying $$\rho_1=U\rho_0 U^{\dag}$$
where $U$ depends on the Hamiltonian of the quantum system, the time $t_0$ and the time $t_1$. When a quantum system couples with its external environment or measurement apparatus, the system becomes an open system and its state evolution is not unitary. For the open quantum system, its evolution can be described by an operator-sum representation in discrete form (see, e.g., \cite{nilsen}) or an appropriate (stochastic) master equation in continuous time (see, e.g., \cite{wisemanbook}).

For a composite quantum system, its state space is the tensor product of the state spaces of its component quantum systems. If we have systems numbered $1$ through $n$ , and the system $j$ ($j=1,\dots,n$) is prepared in the state $\rho_j$, then the joint state of the total system $\rho$ is $$\rho=\rho_1\otimes\rho_2\otimes\cdots\otimes\rho_n$$
where $\otimes$ denotes the tensor product \cite{nilsen}.

\subsection{Quantum CFC}

In quantum CFC, we couple the controlled system (the system plant) to an auxiliary quantum system (the quantum controller), so that the dynamics of the whole system including the system plant and the controller is unitary. Without loss of generality, suppose that the dimension of the whole system consisting of the plant and the controller is finite. Assume that the initial state $\rho_0$ of the whole system is separable and is denoted as $$\rho_0=\rho_0^S\otimes |\psi\rangle_a{_a\langle}\psi|,$$ where $\rho_0^S$ and $|\psi\rangle_a{_a\langle}\psi|$  represent the initial states of the plant and the controller, respectively. The whole system evolves unitarily for a period of time to  $\rho_1$ with
\begin{equation*}\label{cfcunitary}
\rho_1=U (\rho_0^S\otimes|\psi\rangle_a{_a\langle}\psi|) U^{\dag},
\end{equation*}
where the operator $U$ is unitary depending upon the total Hamiltonian of the plant and the controller.

Note that it is the controlled system that we are interested in. In order to obtain the state of the controlled system, the state of the auxiliary system should be traced over from the state of the whole system. According to quantum mechanics, the operation of tracing over the state of the auxiliary system is equivalent to a trace preserving operation on the controlled system \cite{nilsen}.

To be specific, suppose that a set of basis of the Hilbert space of the auxiliary system (with the dimension $d$) is $\{|i\rangle_a\}^{d}_{i=1}$. We define system operators $E_i$ as $E_i={_a\langle} i|U|\psi\rangle_a$ for $i=1,\ \cdots,\ d$. We can verify that $$\sum_{i=1}^{d}E^{\dag}_iE_i=I_S.$$ After one step of quantum CFC, the system state $\rho_1^S$ can be derived by taking the partial trace (Tr$_a$) over $\rho_1$ as follows\footnote{For detailed calculation of partial trace, see, e.g., \cite{nilsen}}:
\begin{equation*}\label{partialtrace}
\rho_1^S=\text{Tr}_a(\rho_1)=\sum_{i=1}^{d} E_i \rho^S_0 E_i^{\dag}.
\end{equation*}

A quantum CFC protocol can involve repeated use of the above process while different auxiliary systems may be used in different steps. Therefore, for a given quantum CFC protocol, there exist generalized operators $\{E_{k,i}:\ i=1,\ \cdots,\ d_k\}$ satisfying $$\sum_{i=1}^{d_k} E_{k,i}^{\dag}E_{k,i}=I_S$$ for all step $k$, such that
\begin{equation}\label{CFC}
\rho_{k+1}^S=\sum_{i=1}^{d_k}E_{k,i}\rho_k^SE_{k,i}^{\dag},
\end{equation}
where $\rho^S_k$ denotes the state of the controlled system at the $k$-th step in the quantum CFC protocol, and $d_k$ is the dimension of the auxiliary system at the $k$-th step.

\subsection{Quantum MFC}

In quantum MFC, we first assume that an MFC protocol consists of a sequence of discrete steps. In each step, we perform a given measurement operation, and then apply a unitary operation to the system conditioned on the available information from the measurement. In quantum mechanics, a measurement will induce  quantum state collapse for a quantum system \cite{nilsen}. Hence, quantum measurement itself could be regarded as a control means, which is quite different from the relevant situation in classical systems. In the following, we will see that a quantum MFC protocol could be equivalent to an adaptive measurement protocol.

In order to compare quantum MFC with the quantum CFC protocol in (\ref{CFC}), we first specify the quantum MFC protocol in the form of discrete steps.
In quantum mechanics, quantum measurement can be described using a set of generalized measurement operators $\{M_n:\ n=1,\ \cdots,\ N\}$, where $$\sum_{n=1}^N M_n^{\dag}M_n=I.$$ At step $k$, we denote the quantum system state before the measurement as $\rho_k$. When we implement a measurement, the result $n$ occurs with probability $p_k^n=\textmd{Tr}(M^{\dag}_nM_n\rho_k)$ and the post-measurement state is
\begin{equation}\label{state-collapse}
\rho_k^n=\frac{M_n\rho_k M_n^{\dag}}{p_k^n}.
\end{equation}
Now a unitary operator $U(\rho_k^n)$ that depends on $\rho_k^n$ is applied to the system. The system state evolves to
\begin{equation}\label{realstate}
\rho_{k+1}^n=U(\rho_k^n)\rho_k^n U^{\dag}(\rho_k^n)=\frac{U(\rho^n_k)M_n\rho_k M_n^{\dag}U^{\dag}(\rho_k^n)}{p_k^n}.
\end{equation}
Here the superscript $n$ of $\rho^n_{k+1}$ depicts the jump direction of the quantum state from step $k$ to $k+1$. To fully monitor a jump trajectory of the quantum state, we need to record all the measurement outcomes in order.

In the process of quantum MFC, if we do not consider the specific measurement result in each step, the non-selective evolution \cite{wisemanbook} of the quantum system can be  obtained by averaging over all possible state trajectories as
$$\rho_{k+1}=\sum_n U(\rho_k^n)M_n\rho_k M_n^{\dag}U^{\dag}(\rho_k^n).$$
Denoting $M_{k,n}=U(\rho_k^n)M_n$, it is clear that $$\sum_{n=1}^{N}M_{k,n}^{\dag}M_{k,n}=I.$$
Hence, $\{M_{k,n}:\ n=1,\ \cdots,\ N\}$ can be considered as a new set of measurement operators depending upon the state at step $k$.
Therefore, the quantum MFC protocol can be equivalently regarded as an adaptive measurement protocol.

If we only consider the non-selective evolution of the quantum system, a quantum MFC protocol can be specified as designing adaptive measurement operators $\{M_{k,n}:\ n=1,\ \cdots,\ N\}$ at step $k$ such that
\begin{equation}\label{MFC}
\rho_{k+1}=\sum_{n=1}^N M_{k,n}\rho_k M_{k,n}^{\dag}.
\end{equation}

Comparing Eq. (\ref{MFC}) with Eq. (\ref{CFC}), we have the following observation.
\begin{observation}\label{observation}
For a given quantum CFC protocol in (\ref{CFC}), one can always find an equivalent non-selective quantum MFC protocol in (\ref{MFC}) provided that the involved generalized measurements can be realized.
\end{observation}

In \cite{kurt-wang}, a comparison between quantum CFC and quantum MFC has been provided where the authors concluded that quantum CFC beats all quantum MFC for the quantum control problems under their consideration. However, the form of quantum MFC in \cite{kurt-wang} was restricted. Specifically, the feedback control operations rely upon the measurement outcome rather than the (estimated) quantum state. Therefore, there are only a limited number of control operations applied on the controlled system. This limits the capability of quantum MFC accordingly. There is an essential difference between state feedback and output feedback for the performance that they can achieve. An example is presented in Appendix A to illustrate the difference between the two types of feedback.
The example also inspires us that the (estimated) state information should be utilized in quantum MFC since it may achieve better performance than that using  measurement output information directly. Therefore, we should reconsider the relative merits between quantum MFC and quantum CFC when the state information can be utilized in quantum MFC.

It is worth stressing that there is no explicit measurement being involved in quantum CFC. This has been widely considered as the remarkable advantage of quantum CFC over quantum MFC. This is because quantum measurement will in general induce  quantum state collapse in a nondeterministic way. Therefore, one has to deal with the additional but inherent uncertainty during the control process in quantum MFC besides the original uncertainty in the system. However, we will demonstrate in the following section that the measurement induced state collapse may play a positive role in feedback control process so that quantum MFC can achieve better performance than quantum CFC in dealing with uncertainties in the initial state.

\section{Real-time Information in Quantum MFC}\label{Real-time-section}
In this section, we compare the control performance that can be achieved in a specific quantum system by the non-selective quantum MFC and the quantum MFC using real-time information.

\subsection{Real-time Information}
When we make a quantum measurement on a quantum system, the measurement outcome is generally stochastic and the measurement induces quantum state collapse accordingly. Therefore, for a monitored quantum system, different measurement records induce different quantum trajectories, which describe the evolutions of the monitored quantum system. However, if we do not utilize the real-time state information that depicts different trajectories of the quantum system, a non-selective quantum MFC protocol in Eq. (\ref{MFC}) can be obtained. In the non-selective quantum MFC, only the averaged information of the quantum state at the current step is utilized in the following  step.

In order to achieve better control performance, a natural idea is to utilize the real-time information of the quantum state trajectory to design a corresponding feedback control law. Here, our focus is on investigating possible advantage of utilizing real-time information in quantum MFC. The investigation can be proceeded by comparing the different control performance that can be achieved by regulating the non-selective evolution of the quantum system or designing a control law using real-time information. When we design a control law using real-time information for a quantum system, a
stochastic master equation (SME) model can be used to describe the evolution of the system dynamics.

The SME model is the continuous-time counterpart of the quantum MFC model shown in Eq. (\ref{state-collapse}) and Eq. (\ref{realstate}) in the sense that it depicts the evolution of a continuously monitored quantum system. The correspondence will be explained explicitly in Subsection IV.B. The stochastic measurement backaction effect is described by a diffusion term in the SME. By the SME, we can obtain the real-time (estimated) information of the quantum state. Actually, it is in essence a filtering equation, i.e.,  an evolution equation of the state estimate of the quantum system driven by the continuous measurement output \cite{hudson,Belavkin1992,quantum-filtering}. If we average the diffusion term in the SME , we can obtain a continuous-time master equation model that corresponds to the  non-selective quantum MFC model in Eq. (\ref{MFC}).

The SME model and the continuous-time master equation describing non-selective evolution have been widely used to investigate open quantum systems (see e.g., \cite{dong2010,wisemanbook}).
The relations between these two models under control have been discussed in \cite{qibo1} by focusing on the unravelling problem. In this paper, we focus on investigating the advantage of using real-time information of the quantum system state in quantum MFC aiming to demonstrate possible different performance that quantum CFC and quantum MFC can achieve.

\subsection{Control Model for Quantum MFC}
We now sketch the model to be used for quantum MFC (see, e.g., \cite{Handel-Stockton2005} for more details). Consider an atomic ensemble consisting of $N$ atoms which are placed into a single mode optical cavity. We consider the ($x, y, z$)-configuration space and assume that the atomic transitions are far detuned from the cavity resonance so that the atomic Hamiltonian can be described by $$H_{A}=\hbar\Delta F_{z}+\hbar u(t)F_{y},$$ where $\Delta$ is the atomic detuning, $F_z$, $F_y$ are the spin-$N/2$ collective dipole moments of the ensemble, and $u(t)$ is the strength of a magnetic field in the $y$-direction and serves as the control input. To detect the state of the atomic ensemble, a probe laser is injected  into the cavity (along $z$-direction) by a beamsplitter and the optical field is configured to good approximation so that it only interacts with the collective angular momentum degrees of the atoms. After interacting with the atomic ensemble, the outgoing optical field is detected by a Homodyne detection. Based on the flow of measurement output, we can estimate the state of the atomic ensemble.

The SME describing the conditional evolution of the atomic state $\rho_c(t)$ driven by the measurement output is \cite{Handel-Stockton2005}
\begin{align}\label{cl1}
&\ d\rho_c(t)=-i u(t)[F_{y} \ , \rho_c(t)] dt-is[F_{z}
,\rho_c(t)]dt\nonumber\\  &\ \ \ \ \ \
\ \ \ \ \ +M\mathcal{D}[F_{z}]\rho_c(t) dt+\sqrt[]{M\eta} \mathcal{H}[F_{z}]\rho_c(t)dW_{t},
\end{align}
where  $s$ is related to the experimental parameters such as $\Delta$ and so on, $M$ is the measurement rate, $\eta$ is the detection efficiency, and
$$\mathcal{D}[\Lambda]\rho=\Lambda\rho\Lambda^{\dag}-\frac{1}{2}(\Lambda^\dag\Lambda\rho+\rho\Lambda^\dag\Lambda),$$
$$ \mathcal{H}[\Lambda]\rho=\Lambda\rho +\rho \Lambda^{\dag}
-\mathrm{Tr}(\Lambda\rho+\rho\Lambda^{\dag})\rho.$$
The innovation process $W_t$ is a Wiener process satisfying
\begin{equation}\label{w}
\sqrt{\eta}dW_{t}=dy_{t}-2\ \sqrt{M}\eta\ \mathrm{Tr}\ (F_{z}\rho_c(t))\ dt,
\end{equation}
where $y_t$ is the flow of measurement output.

It is worth pointing out that the whole measurement backaction effect, i.e., the deterministic drift part $M\mathcal{D}[F_{z}]\rho_c(t) dt$ as well as the diffusion part $\sqrt{M\eta} \mathcal{H}[F_{z}]\rho_c(t)dW_{t}$, is fully included in the SME model. The diffusion part in Eq. (\ref{cl1}) can play an important role when utilizing the real-time information for feedback control.

Now we demonstrate that the SME in Eq. (\ref{cl1}) can correspond to the discrete model in Eq. (\ref{state-collapse}) and in Eq. (\ref{realstate}) since Eq. (\ref{cl1}) can be obtained starting from a form as Eq. (\ref{state-collapse}). Actually, in the discrete-time model (\ref{state-collapse}), we assume that there are only finite number of measurement outcomes for each step. However, the measurement operator may have a continuous spectrum. As in the above model, it is the field quadrature being continuously measured under the Homodyne detection. The infinitesimal measurement output $dy_t$ has a Gaussian distribution with mean $2\sqrt{M}\eta\langle F_z\rangle dt$ and variance $\eta dt$, where $$\langle \cdot \rangle=\mathrm{Tr}(\cdot \rho_c(t)).$$ Hence, $dy_t$ can be expressed as
$$dy_t=2\sqrt{M}\eta\langle F_z\rangle dt+\sqrt{\eta}dW_t.$$
According to quantum mechanics, when the measurement outcome $dy_t$ occurs, the conditional quantum state evolves as
$$\rho_c(t+dt)=\frac{\Omega(dy_t)\rho_c(t)\Omega^{\dag}(dy_t)}{\mathrm{Tr}
[\Omega(dy_t)\rho_c(t)\Omega^{\dag}(dy_t)]},$$
where the (unnormalized) measurement operator $$\Omega(dy_t)=I-i[u(t)F_y+sF_z]dt-\frac{M}{2}F^2_zdt+dy_t\sqrt{M}F_z.$$
Moreover, we have
\begin{align*}
\rho_c(t+dt)&=\frac{\Omega(dy_t)\rho_c(t)\Omega^{\dag}(dy_t)}{\mathrm{Tr}
[\Omega(dy_t)\rho_c(t)\Omega^{\dag}(dy_t)]}\\
&=\frac{\rho_c(t)-i[u(t)F_y+sF_z, \rho_c(t)]dt+\mathcal{D}[\sqrt{M}F_z]\rho_c(t)dt}{1+2\sqrt{M}\langle
F_z\rangle dy_t}+\frac{\sqrt{M}[F_z\rho_c(t)+\rho_c(t)F_z]dy_t}{1+2\sqrt{M}\langle
F_z\rangle dy_t}\\
&=\{\rho_c(t)-i[u(t)F_y+sF_z, \rho_c(t)]dt+\mathcal{D}[\sqrt{M}F_z]\rho_c(t)dt\\
&~~~+\sqrt{M}[F_z\rho_c(t)+\rho_c(t)F_z]dy_t\}\{1-2\sqrt{M}\langle F_z\rangle dy_t+4\eta\langle F_z\rangle^2dt\}\\
&=\rho_c(t)-i[u(t)F_y+sF_z, \rho_c(t)]dt+\mathcal{D}[\sqrt{M}F_z]\rho_c(t)dt\\
&~~~+\mathcal{H}[\sqrt{M}F_z]\rho_c(t)[dy_t-2\sqrt{M}\eta\langle F_z\rangle dt]\\
&=\rho_c(t)-i[u(t)F_y+sF_z, \rho_c(t)]dt+M\mathcal{D}[F_z]\rho_c(t)dt+\sqrt{M\eta}\mathcal{H} [F_z]\rho_c(t)dW_t.
\end{align*}
Hence, the SME (\ref{cl1}) is obtained.

Now we turn to the continuous-time non-selective model for quantum MFC. The situation can correspond to the case where either we cannot obtain the measurement output or our concern is only the average evolution of the quantum state. The model can be derived by averaging the conditional evolution of $\rho_c(t)$ and it can be described as
\begin{equation}\label{me}
\frac{d \rho_{t}}{dt}=-i [s F_{z}+u(t)F_{y} ,\rho_{t}]
+M\mathcal{D}[F_{z}] \rho_{t}.
\end{equation}
Note that in contrast to the SME model (\ref{cl1}), only the drift part $M\mathcal{D}[F_{z}] \rho_{t}dt$ of the measurement backaction effect is retained in (\ref{me}). The model can be regarded as a continuous-time counterpart of Eq. (4).

\subsection{Performance Comparison}
Suppose that the initial state of the system is $\rho_{0}=\sum^{n}_{i=1} p_{i}\rho_{i}$, where the state $\rho_{i}$ has the corresponding probability $p_{i}$, $\sum_{i=1}^{n}p_{i}=1$, $n\geq 2$. Note that if the initial state is mixed, its entropy is not zero. In this sense, we say that there is some uncertainty in the initial state. The control objective is to prepare an arbitrarily desired eigenstate of $F_{z}$ with a high level of fidelity from an arbitrary initial state. We have the following result.

\begin{proposition}\label{2}
Using the non-selective quantum MFC in (\ref{me}), one cannot prepare an arbitrarily desired eigenstate of $F_z$ from a mixed initial state no matter how to design the control law $u_t$. The state can be globally stabilized to a desired eigenstate of $F_z$ if the quantum MFC using real-time information in (\ref{cl1}) is employed.
\end{proposition}
\textbf{Proof.}
When we employ the non-selective quantum MFC in (\ref{me}) for feedback control,
let us first consider the dynamics of
$\textmd{Tr}(\rho^2_{t})$. From (\ref{me}), we have
\begin{align*}
\frac{d\textmd{Tr}(\rho_{t}^{2})}{dt}
&=2\textmd{Tr}[-i(sF_z+u_{t}F_y)\rho_{t}^{2}+i\rho_{t}^{2}(sF_z+u_{t}F_y)+MF_z\rho_{t}F_z\rho_{t}-\frac{M}{2}(F^2_z\rho_{t}^{2}+\rho_{t}^{2}F^2_z)]\\
&=2M \textmd{Tr}[F_z\rho_{t}F_z\rho_{t}-F^2_z\rho_{t}^{2}]\\
&\leq 2M\textmd{Tr}[F_z\rho_t^2F_z-F^2_z\rho_{t}^{2}]\\
&=0.
\end{align*}
Hence, $\frac{d\textmd{Tr}(\rho_{t}^{2})}{dt}\leq 0,$ which implies that
for all $t\geq 0$, $$\textmd{Tr}(\rho^2_t)\leq\textmd{Tr}(\rho^2_0).$$
Note that $\textmd{Tr}(\rho^2)=1$ if and only if $\rho$ is a pure state. Therefore,
we conclude that the desired target state cannot be prepared from
a mixed initial state $\rho_0$. Hence, we cannot prepare an arbitrarily desired eigenstate of $F_z$ from a mixed initial state no matter how to design the feedback control law $u(\cdot)$.

When the real-time estimated state information $\rho_c(t)$ may be utilized for quantum feedback control, we can resort to Theorem $4.2$ of Ref. \cite{Mirrahimi-Handel-2007}, which has been restated as follows for completeness.
\begin{theorem}\label{1}
\cite{Mirrahimi-Handel-2007} For the SME model (\ref{cl1}), denote an arbitrary desired eigenstate of $F_z$ as $\rho_d$. Consider the following control law:
\begin{description}
  \item[1.]If $\textmd{Tr}(\rho_{c}(t)\rho_{d})\geq\gamma$, $u_{t}=-\textmd{Tr}(i[F_y,\rho_{c}(t)]\rho_{d})$;
  \item[2.]If $\textmd{Tr}(\rho_{c}(t)\rho_{d})\leq\gamma/2$, $u_{t}=1$;
  \item[3.]If $\rho_{c}(t)\in\mathcal{B}\triangleq\{\rho:~\gamma/2<\textmd{Tr}(\rho\rho_{d})<\gamma\}$, then $u_t=-\textmd{Tr}(i[F_y,\rho_{c}(t)]\rho_{d})$ if $\rho_{c}(t)$ last entered into $\mathcal{B}$ through the boundary $\textmd{Tr}(\rho\rho_{d})=\gamma$; $u_t=1$ otherwise.
\end{description}
Then there exists $\gamma>0$ such that $u_t$ globally stabilizes (\ref{cl1}) around $\rho_d$ and $E\rho_c(t)\rightarrow \rho_d$ as $t\rightarrow \infty$.
\end{theorem}

Theorem \ref{1} proposed an explicit control law which can globally stabilize (\ref{cl1}) around an arbitrary desired eigenstate $\rho_{d}$ of $F_z$. That is, we can approximately prepare the target state from an arbitrary initial state with probability 1.
\ \hfill$\Box$

Proposition 2 provides an example that the quantum MFC using real-time information can achieve better performance than the non-selective quantum MFC for some practical tasks. In the SME model (5), the whole measurement backaction effect (i.e., the drift part $M\mathcal{D}[F_{z}]\rho_c(t) dt$ as well as the diffusion part $\sqrt{M\eta} \mathcal{H}[F_{z}]\rho_c(t)dW_{t}$) is fully included  and the feedback control law is based on the real-time state information updated by the measurement output flow. In the non-selective quantum MFC in (7), only the averaged state information can be utilized. The different effects can be explicitly illustrated by considering the simplest case, i.e., $N=1$.

{\bf Example 2.}
Denote ~$|0\rangle=\begin{pmatrix} 1\\
0\end{pmatrix}$ and $|1\rangle=\begin{pmatrix} 0\\ 1\end{pmatrix}$
as the two eigenvectors of ~$F_{z}$. Under this vector
representation, we have
$$F_{z}=\frac{1}{2}(|1\rangle \langle 1|-|0\rangle \langle
0|)=-\frac{1}{2}\begin{pmatrix}1&0\\0&-1\end{pmatrix},$$
$$F_{y}=\frac{1}{2}(i|0\rangle \langle 1|-i|1\rangle \langle
0|)=-\frac{1}{2}\begin{pmatrix}0&-i\\i&0\end{pmatrix}.$$

For the two-level system, we can use Bloch vector $\mathbf{r}=(x,y,z)$ to represent the quantum state $\rho$ as
$\rho=\frac{1}{2}\begin{pmatrix}1+z&x-iy\\x+iy&1-z\end{pmatrix}$. A brief introduction to Bloch representation is presented in Appendix B. With the Bloch representation, the SME (\ref{cl1}) can be converted into
\begin{equation}\label{condf}
\begin{aligned}
dx_{c}(t)&=-\frac{M}{2}x_{c}(t)dt-u_{t}z_{c}(t)dt +sy_{c}(t)dt+\sqrt{M\eta}x_{c}(t)z_{c}(t)dW_{t} \\
dy_{c}(t)&=-\frac{M}{2}y_{c}(t)dt-sx_{c}(t)dt+\sqrt{M\eta}y_{c}(t)z_{c}(t)dW_{t}\\
dz_{c}(t)&=u_{t} x_{c}(t)dt-\sqrt{M\eta}(1-z^{2}_{c}(t))dW_{t}.
\end{aligned}
\end{equation}
The non-selective quantum MFC model in (\ref{me}) becomes
\begin{equation}\label{masfen}
\begin{aligned}
\frac{dx_{t}}{dt}&=-\frac{M} {2}x_{t}-u_{t} z_{t}+s y_{t}\\
\frac{dy_{t}}{dt}&=-\frac{M}{2}y_{t}-s x_{t}\\
\frac{dz_{t}}{dt}&=u_{t} x_{t}.
\end{aligned}
\end{equation}

It is easy to see that $\left(x, y, z \right)=\left(0, 0, 0\right)$ is an equilibrium point of (\ref{masfen}). Thus, once $\rho=\frac{I}{2}$ during the control process, it will be stuck at
this point no matter  what control law is performed. Hence, we cannot prepare the target state (corresponding to $z$ = 1 or -1) from an arbitrary initial state by using the non-selective quantum MFC model (\ref{masfen}). In contrast,  the diffusion term depicting the stochastic measurement
backaction effect in the SME model (\ref{condf}) can  avoid the occurrence of this kind of pitfalls. Thus, we can utilize the real-time state information to design a corresponding feedback control law to regulate the state trajectory to achieve the target state.

The difference behind the SME model (\ref{condf}) and the non-selective quantum MFC model in (\ref{masfen}) can also be demonstrated by considering the evolution of $\widetilde{\rho}(t)=E\rho_{c}(t)$. From (\ref{condf}), we have
\begin{equation}\label{ensfen}
\begin{aligned}
\frac{d\widetilde{x}(t)}{dt}&=-\frac{M} {2}\widetilde{x}(t)-Eu_{t}\widetilde{z}_{c}(t)+s \widetilde{y}(t)\\
\frac{d\widetilde{y}(t)}{dt}&=-\frac{M}{2}\widetilde{y}(t)-s \widetilde{x}(t)\\
\frac{d\widetilde{z}(t)}{dt}&=Eu_{t}\widetilde{x}_{c}(t).
\end{aligned}
\end{equation}
Note that $Eu_{t}\widetilde{z}_{c}(t)\neq \widetilde{z}(t)Eu_{t}$ and
$Eu_{t}\widetilde{x}_{c}(t)\neq \widetilde{x}(t)Eu_{t}$ in general since $u_t$ may be a function of
$x_{c},\ y_{c}$ and $z_{c}$. In the non-selective quantum MFC model in (\ref{masfen}), for each infinitesimal step, the real-time state information is first averaged and then the averaged information is used for feedback control. However, in (\ref{ensfen}), we utilize the state trajectory information for feedback control and then an ensemble average is made. The essential difference results in different control performance that they can be achieved.

Example 2 shows that it is the real-time state information as well as the stochastic measurement backaction effect that helps the feedback control law achieve the control objective for the quantum MFC using the SME model (\ref{cl1}). The advantage of using the real-time state information in quantum feedback control can be explicitly demonstrated for the specific practical task.

Recall that the quantum CFC protocol (\ref{CFC}) also uses the averaged state information for feedback control and it can be equivalently regarded as a non-selective quantum MFC protocol. Therefore, quantum MFC using real-time state information can achieve better performance than quantum CFC theoretically.

\section{Hamiltonian Parameter Uncertainty in Quantum Feedback Control}

When we focused on quantum MFC using real-time information in Section IV, we assumed no uncertainty in the system Hamiltonian. For a practical quantum system, the existence of uncertainties in the system Hamiltonian is unavoidable (e.g., due to imprecise model and control error)  \cite{dong-petersen1,dong-petersen2}. As we mentioned in Section \ref{second} the advantage of feedback can be taken when we utilize real-time information to deal with various uncertainties. In this section, we compare the different performance that quantum MFC and quantum CFC can achieve in dealing with the Hamiltonian parameter uncertainty by an example of quantum state preparation.

The objective is to prepare a specified pure state from an arbitrary initial state for a quantum system. Suppose that the dimension of the state space of the quantum system  is $d$. For a specified pure state $|\psi\rangle$, without loss of generality, one can always find a set of orthonormal basis $\{|0\rangle, \cdots, |d-1\rangle\}$ such that the target state $|\psi\rangle=|d-1\rangle$.

The task can be easily fulfilled provided that the set of generalized measurement operators $$\{E_i=|d-1\rangle\langle i|:\ i=0,\ \cdots,\ d-1\}$$ can be realized. Actually, we can verify that $$\sum_{i=0}^{d-1} E^{\dag}_i E_i=I,$$ and for an arbitrary initial state $\rho$, $$\sum_{i=0}^{d-1} E_i\rho E_i^{\dag}=|d-1\rangle\langle d-1|=|\psi\rangle \langle\psi|.$$

To compare the different performance that quantum MFC and quantum CFC can achieve when dealing with Hamiltonian parameter uncertainty, we present the following example with $d=2$ and $|\psi\rangle=|1\rangle$.

{\bf Example 3.}
The total Hamiltonian of the plant and its auxiliary system is
\begin{equation}\label{hamiltonian}
H=i\theta (a^{\dag}b-ab^{\dag}),\footnote{We have set $\hbar=1$.}
\end{equation}
where $b$ and $a$ denote the annihilation operators on the plant and auxiliary system, respectively, and $\theta$ depicts the coupling between the plant and auxiliary system.  With this  Hamiltonian, the total unitary evolution is
\begin{equation}\label{unitary}
U_t=\textmd{exp}[(a^{\dag}b-ab^{\dag})\theta t].
\end{equation}
It can be verified that
\begin{align}\label{operations}
   U_t|0\rangle_b|0\rangle_a&=|0\rangle_b|0\rangle_a\\ \nonumber
  U_t|1\rangle_b|0\rangle_a&=\textmd{cos}(\theta t)|1\rangle_b|0\rangle_a-\textmd{sin}(\theta t)|0\rangle_b|1\rangle_a\\ \nonumber
   U_t|0\rangle_b|1\rangle_a&=\textmd{sin}(\theta t)|1\rangle_b|0\rangle_a+\textmd{cos}(\theta t)|0\rangle_b|1\rangle_a\\ \nonumber
   U_t|1\rangle_b|1\rangle_a&=\textmd{cos}(2\theta t)|1\rangle_b|1\rangle_a+\frac{\sqrt{2}}{2}\textmd{sin}(2\theta t)(|2\rangle_b|0\rangle_a-|0\rangle_b|2\rangle_a).
\end{align}

Assume that the objective is to prepare $|1\rangle_b$ from an arbitrary initial state. When the Hamiltonian parameter $\theta$ is known accurately, if we set the state of the auxiliary system being $|1\rangle_a$, and  $t=\frac{\pi}{2\theta}$, then $$E_0={}_a\langle0|U_{\frac{\pi}{2\theta}}|1\rangle_a=|1\rangle_b{_b\langle}0|,\
E_1={}_a\langle1|U_{\frac{\pi}{2\theta}}|1\rangle_a|=|1\rangle_b{_b\langle}1|.$$ From Eq. (\ref{CFC})  and Eq. (\ref{MFC}), we can conclude that if there is no uncertainty with the total Hamiltonian, we can achieve the control target either using quantum CFC or quantum MFC.

Now we suppose that the parameter $\theta$ in the total Hamiltonian of the plant and the auxiliary system is unknown or our knowledge of $\theta$ is not accurate.
Once there is parameter uncertainty in the total Hamiltonian, it will generally result in parameter uncertainty in the generalized measurement operators $\{E_{t,i}\}$ in Eq. (\ref{CFC}) accordingly when the quantum CFC protocol  (\ref{CFC}) is used. In the quantum CFC protocol, there is no explicit measurement during the whole control process. Hence, one cannot reduce this kind of parameter uncertainty  and cannot achieve the control target in general.  Specifically, once $\theta$ is unknown in Example 3, one has no ways to identify it. Therefore, one cannot identify the time duration for quantum CFC to fulfill the control task.

When we employ the quantum MFC protocol, for every copy of the system, we can couple an auxiliary system initialized in $|1\rangle_a$, and let the total system evolve for a period of short time. After that we can perform a projective measurement on the auxiliary system and choose the subsequent operations based on the real-time measurement outcome. In particular, no matter what the initial state of the system is in Example 3, we can first perform a projective measurement on the system. The post-measurement state is either $|1\rangle_b$ or $|0\rangle_b$.  If it is $|1\rangle_b$, the task is accomplished. While if it is $|0\rangle_b$, we can couple the plant and the auxiliary system initialized in $|1\rangle_a$ for a short time duration $t$  followed by a projective measurement on the auxiliary system. From (\ref{operations}), the total system will be in $|0\rangle_b|1\rangle_a$  ($|1\rangle_b|0\rangle_a$) with probability cos$^2$$(\theta t$) (sin$^2$($\theta t$)). If the auxiliary state is $|0\rangle_a$, the system state has been prepared in $|1\rangle_b$. If the auxiliary state is $|1\rangle_a$, we can repeat the above process, i.e., coupling the plant and  the auxiliary system in $|1\rangle_a$ for a short time duration $t$ followed by a projective measurement on the auxiliary system.  After $n$ times of repetitions, the probability of the system state still being $|0\rangle_b$ is cos$^{2n}(\theta t)$, which approaches to zero exponentially with $n$. Thus, we can utilize the real-time information about the quantum state of the system to achieve the control target using quantum MFC although $\theta$ is unkown.

If our knowledge of  $\theta$ is not accurate, i.e., $\bar{\theta}\neq \theta$, the quantum MFC protocol can still be applied. Moreover, if decoherence occurs with the system (e.g., the system state $|1\rangle_b$ decays to $|0\rangle_b$), we can identify whether the decoherence happens, and correct it if necessary using the above quantum MFC protocol.

In addition to the above quantum MFC protocol, there is another method to achieve the objective in the presence of Hamiltonian parameter uncertainty. We can first estimate the unknown parameter and then design the control law based on the obtained estimate. How to precisely estimate unknown parameters is an extremely important objective in quantum information and quantum metrology \cite{berry,armen,wiseman-berry1,qi-hou,mahler,peaudecerf}. Recent results have demonstrated that adaptive quantum measurement can greatly improve the estimation accuracy \cite{berry,armen,wiseman-berry1,mahler,peaudecerf}.  The estimate process involving adaptive quantum measurement can be regarded as a specific quantum MFC. With an accurate  estimate, we can design a feedback control law to prepare the target state with a high level of fidelity. Hence, quantum MFC can be superior to quantum CFC in dealing with Hamiltonian parameter uncertainty when manipulating some specific quantum systems.

\section{Concluding Remarks}
In this paper, we compare quantum coherent feedback control (CFC) and quantum measurement-based feedback control (MFC) by several examples. Based on an equivalent relationship between quantum CFC and non-selective quantum MFC in the form of operator-sum representation, we show that quantum MFC can achieve better performance than quantum CFC in stabilizing a quantum state and dealing with Hamiltonian parameter uncertainty. The use of real-time information for feedback control design in quantum MFC is a key factor that shows the advantage of quantum MFC over quantum CFC.

In the analysis, we assume that all of the generalized measurements necessary for quantum MFC should be able to be realized and no time delay exists in the feedback loop. For a practical quantum system, these assumptions may not be guaranteed. For example, time delay is one of the major difficulties in implementing quantum MFC on a quantum system since it always takes time to acquire real-time information and to update the conditional state for feedback control design. Although the effect of time delay may be reduced for some specific tasks \cite{amini}, the effect of its existence will degrade the performance that ideal quantum MFC can achieve in general. In contrast, it may not be necessary to worry about the time delay in the feedback loop for quantum CFC since the controlled system and the controller have similar time scales and no measurements are required in the feedback design. Quantum CFC can show significant advantages over quantum MFC in some aspects such as practical implementation and high speed. The results in this paper show that quantum MFC can theoretically achieve better performance than quantum CFC in dealing with uncertainties in the initial state and Hamiltonian. With the development of real-time quantum MFC experiments (see e.g., \cite{sayrin,vijay}), the results may provide useful information in choosing suitable feedback protocols for manipulating quantum systems. The work may also enrich understanding of the relative advantages between quantum CFC and quantum MFC, and promote the investigation on the capability of feedback in dealing with uncertainties for quantum systems.

\section*{Acknowledgements}
The authors thank Lei Guo and Li Li for helpful discussions.

\section*{Appendices}
\subsection*{Appendix A: Output feedback vs state feedback}
We demonstrate the difference between output feedback control and state feedback control by the following example.

{\bf Example 4.} Consider the linear system
\begin{equation*}
\dot{x}=Ax+Bu,\     \   y=Cx,
\end{equation*}
where $x$ is the system state, $u$ is the control input, $y$ is the measurement output, and $$A=\left(
           \begin{array}{cc}
             0 & 1 \\
             -1 & 0 \\
           \end{array}
         \right)
,\ B=\left(
        \begin{array}{c}
          0 \\
          1 \\
        \end{array}
      \right)
,\ C=\left(
        \begin{array}{cc}
          1 & 0 \\
        \end{array}
      \right).$$
It is easy to verify that the linear system is controllable and observable \cite{linearsystem}.
Assume that the objective is to stabilize the system state to $x=0$ asymptotically.

If we use the state information for feedback, i.e., the control law is in the form of $u=Kx$, the control objective can be achieved with $K=\left(
     \begin{array}{cc}
       -1 & -3 \\
     \end{array}
   \right)
$.
However, if the control law is in the form of $u=Ly$, i.e., the measurement output information $y$ is used, one can prove that the control objective cannot be achieved no matter how to design $L$, even though the system is controllable.

One may argue that we only have access to the measurement output of the system so that we cannot use the state feedback control protocol unless the system state can be observed. Note that the linear system is controllable and observable. We can  construct a state observer (state estimate) $z(t)$ based on the measurement output information, and utilize the estimated state information $z(t)$ for feedback control.

In the example, we can construct $z(t)$  as
\begin{equation*}
\dot{z}=(A-GC)z+Gy+Bu,
\end{equation*}
where $G=\left(
           \begin{array}{c}
             3 \\
             1 \\
           \end{array}
         \right)
$, and $u=Kz$.
The whole system consisting of the plant and the state observer evolves as
\begin{equation*}
\left(
  \begin{array}{c}
    \dot{x} \\
    \dot{z} \\
  \end{array}
\right)=\left(
          \begin{array}{cc}
            A & BK \\
            GC & A-GC+BK \\
          \end{array}
        \right)\left(
                 \begin{array}{c}
                   x \\
                   z \\
                 \end{array}
               \right).\\
               \end{equation*}
Since all the eigenvalues of $$\left(
          \begin{array}{cc}
            A & BK \\
            GC & A-GC+BK \\
          \end{array}
        \right)=\left(
                  \begin{array}{cccc}
                    0 & 1 & 0 & 0 \\
                    -1 & 0 & -1 & -3 \\
                    3 & 0 & -3 & 1 \\
                    1 & 0 & -3 & -3 \\
                  \end{array}
                \right)
        $$ are real and negative, the control objective can be achieved.

\subsection*{Appendix B: Bloch representation}
For a two-level
quantum system, the state $\rho$ can be represented in terms of the
Bloch vector
$\mathbf{r}=(x,y,z)$.
If we denote the Pauli
matrices $\sigma=(\sigma_{x},\sigma_{y},\sigma_{z})$ as follows:
\begin{equation}
\sigma_{x}=\begin{pmatrix}
  0 & 1  \\
  1 & 0  \\
\end{pmatrix} , \ \ \ \
\sigma_{y}=\begin{pmatrix}
  0 & -i  \\
  i & 0  \\
\end{pmatrix} , \ \ \ \
\sigma_{z}=\begin{pmatrix}
  1 & 0  \\
  0 & -1  \\
\end{pmatrix} ,
\end{equation}
the Bloch vector $\mathbf{r}=(x,y,z)$ can be calculated as
$x=\text{Tr}(\rho\sigma_{x})$, $y=\text{Tr}(\rho\sigma_{y})$ and $z=\text{Tr}(\rho\sigma_{z})$.
The density matrix $\rho$ can be represented using the Bloch vector as
\begin{equation}\label{eq4}
\rho=\frac{1}{2}(I+\mathbf{r}\cdot \sigma) .
\end{equation}
Each point on the unit Bloch sphere (i.e., $x^2+y^2+z^2=1$) corresponds to a pure state of the two-level quantum system and each point in the interior of the sphere corresponds to a mixed state.

 \end{document}